\documentclass[12pt]{article}
\usepackage{amssymb,amsmath,hyperref}

\def\lromn#1{\uppercase\expandafter{\romannumeral#1}}

\usepackage{graphicx}

\newcommand{\gsim}{ \mathop{}_{\textstyle \sim}^{\textstyle >} }
\newcommand{\lsim}{ \mathop{}_{\textstyle \sim}^{\textstyle <} }

\begin{document}

\begin{titlepage}

\begin{center}

\hfill ICRR-Report-522-2005-5\\
\hfill KEK-TH-1047\\
\hfill \today

\vspace{1cm}
{\large Heavy Wino-like Neutralino Dark Matter Annihilation\\ into Antiparticles}
\vspace{1cm}

{\bf Junji Hisano}$^{1}$,
{\bf Shigeki Matsumoto}$^{2}$,
{\bf Osamu Saito}$^{1}$,
and
{\bf Masato Senami}$^{1}$
\vskip 0.15in
{\it
$^1${ICRR, University of Tokyo, Kashiwa 277-8582, Japan}\\
$^2${Theory Group, KEK, Oho 1-1 Tsukuba, Ibaraki 305-0801, Japan}
}
\vskip 2in

\abstract{The lightest neutralino is a viable dark matter (DM)
  candidate. In this paper we study indirect detection of the
  wino-like neutralino DM using positrons and antiprotons from the
  annihilation in the galactic halo. When the mass is around 2~TeV,
  which is favored from the thermal relic abundance, the
  non-perturbation effect significantly enhances the annihilation
  cross sections into positrons and antiprotons.  We find that the
  positron and antiproton fluxes with energies larger than 100~GeV may
  become larger than the expected backgrounds. Since the positron flux
  is less sensitive to the astrophysical parameters, the detection may
  be promising in the upcoming experiments such as PAMELA and AMS-02.
  We also find the wino-like neutralino DM with mass around 2~TeV is
  compatible with the HEAT anomaly. }

\end{center}
\end{titlepage}
\setcounter{footnote}{0}

\section{Introduction}

The existence of the cold dark matter (CDM) has been confirmed by the
WMAP measurement of the cosmic microwave background
\cite{Spergel:2003cb}; $\Omega_{{\rm CDM}}h^2=0.113^{+0.016}_{-0.018}$
\cite{Bennett:2003bz}.  However, the nature of the dark matter (DM)
still remains a mystery. Weakly Interacting Massive Particles (WIMPs)
are viable candidates for the DM since their thermal relic abundances
are naturally within the observed range \cite{Jungman:1995df}.

A well-studied representative of WIMPs is the lightest neutralino in
supersymmetric extensions of the standard model \cite{Nilles:1983ge}.
Neutralinos are composed of bino, neutral wino and neutral Higgsinos,
which are superpartners of the U(1)$_Y$ and SU(2)$_L$ gauge bosons and
the Higgs bosons, respectively. In most supersymmetric models, the
lightest neutralino is the lightest supersymmetric particle (LSP) and
is stable due to the R-parity conservation. The constituent of the
neutralino depends on supersymmetry breaking models. For example, the
neutralino is bino-like in a wide region of the parameter region of
the minimal supergravity model. In the anomaly mediated supersymmetry
breaking model (AMSB) \cite{Randall:1998uk}, the neutralino is
wino-like because gaugino masses are proportional to beta functions of
the gauge coupling constants. The wino-like neutralino has larger
coupling than the bino-like one, so that the wino-like neutralino DM
has larger prospects for detection.

Various experiments have been performed or are planed in order to
detect the neutralino DM directly or indirectly. The direct detections
are to measure the recoil energy which the neutralino may deposit as it
crosses a terrestrial detector \cite{Goodman:1984dc}. The detection
rate depends on the cross section for the elastic-scattering of the
neutralino with target nuclei. On the other hand, the indirect ones
are to detect the anomalous cosmic rays produced in the neutralino
annihilation. Detectors are designed to observe high-energy neutrinos
from the earth or the sun, gamma-rays from the galactic center, and
antimatter cosmic rays from the galactic halo, which are generated
from the neutralino annihilation.

In this paper, we consider indirect detection of the wino-like
neutralino DM by positrons and antiprotons in cosmic rays.  It is
pointed out in Refs.~\cite{Hisano:2003ec,Hisano:2004ds} that the cross
sections for the wino-like neutralino annihilation into gauge
bosons are enhanced compared with those at the tree-level
approximation when the mass is larger than about 1~TeV.  This is due to a
non-perturbative effect by the electroweak interaction, which appears
in a non-relativistic limit of the wino-like neutralinos.
Especially, when the mass is around 2~TeV, which is favored from
the thermal relic abundance of the wino-like neutralino
\cite{Profumo:2004at}, the annihilation cross sections are
significantly enhanced by the resonance effect.

The enhancement of the annihilation cross sections raises the
possibilities of the indirect detection of the wino-like neutralino
DM.  In Ref.~\cite{Hisano:2004ds}, the gamma-ray flux produced by the
wino-like neutralino annihilation in the galactic center is evaluated,
and it is found that the the sensitivity for the wino-like neutralino
DM is enhanced. In this paper, we evaluate the positron and 
antiproton fluxes from the wino-like neutralino annihilation in the
galactic halo, including the non-perturbative effect. We find that,
for the neutralino with mass around 2~TeV, the positron and antiproton
signals also exceed the backgrounds.  These fluxes will be measured
with unprecedented accuracies by the upcoming experiments such as
PAMELA \cite{Boezio:2004jx} and AMS-02 \cite{Barao:2004ik}.
Especially, the measurement of the positron flux may be more promising
for detection of the wino-like neutralino with mass around 2~TeV,
since the predicted positron flux is less sensitive to the
astrophysical parameters responsible to the propagation or the DM halo
profile.

This paper is organized as follows. In section 2, we review the
non-perturbative effect on the wino-like neutralino annihilation cross
sections. In section 3, the positron flux from the annihilation in the
galactic halo is evaluated using the diffusion model. Here, we compare
the predicted signal positron flux with the expected background, and
discuss the sensitivities of the future experiments to the heavy
wino-like neutralino DM.  The HEAT anomaly \cite{Barwick:1997ig} is
also discussed. In section 4, we investigate the antiproton flux from
the wino-like neutralino annihilation. The expected background and the
future prospect are also discussed.  Section 5 is devoted to
conclusions.

\section{Non-perturbative effect on wino-like neutralino annihilation}
\label{sec:resonance}

The wino-like neutralinos annihilate mainly into $W$ bosons due to the
SU(2)$_L$ gauge interaction.  The annihilation process is mediated by
$t$-channel wino-like chargino exchange at tree-level, and the cross
section is given by
\begin{equation}
  \sigma v =
  \frac{2\pi\alpha_2^2}{m^2}~,
\label{tree}
\end{equation}
where $v$ is the relative velocity of the neutralinos,
$\alpha_2$ is the SU(2)$_L$ gauge coupling constant and $m$ is the wino
mass.  

We take a non-relativistic limit ($v\ll 1$) in Eq.~(\ref{tree}),
however, the tree-level approximation in the limit is not valid for
the wino-like neutralino heavier than $\sim m_W/\alpha_2$. Here,
$m_W$ is the $W$ boson mass. This is due to the threshold singularity
caused by the mass degeneracy between the wino-like neutralino and
chargino, and the higher-order contributions should be included in the
case. The dominant contribution to the scattering amplitude at
$O(\alpha_2^{(n+1)})$ comes from ladder diagrams, in which $n$ gauge bosons
are exchanged. When the mass difference between the wino-like
neutralino and chargino is negligible, the $n$-th ladder diagram is
suppressed by only $(\alpha_2 m/m_W)^{(n)}$ compared with the
leading-order one \cite{Hisano:2002fk}. Thus, when $m$ is larger than
$\sim m_W/\alpha_2$, we need to resum the diagrams at all orders. In
other words, we have to include the non-perturbative effect for
obtaining the reliable annihilation cross section.

The resummation of the ladder diagrams has a following
interpretation. Since the wino mass is much heavier than that of the
$W$ boson, the wino-like neutralinos feel the long-range force induced
from the $W$ boson exchange.  Due to the force, the wave function of the
neutralino pair is significantly modified from the plane wave before
the annihilation into $W^+W^-$ bosons. As shown in
Ref.~\cite{Hisano:2004ds}, the bound states, which are composed of the
neutralino and chargino pairs, appear due to the long-range force if
the wino mass is large enough. Especially interesting, a
bound state has the binding energy almost zero when the wino
mass is close to $\sim$ 2, 8, $\cdots$~TeV. In those cases, the
wino-like neutralino annihilation cross section in a non-relativistic
limit is enhanced by several orders of magnitude compared to that of
the tree-level cross section due to the resonance.

\begin{figure}[t]
 \begin{center}
  \scalebox{0.58}{\includegraphics*{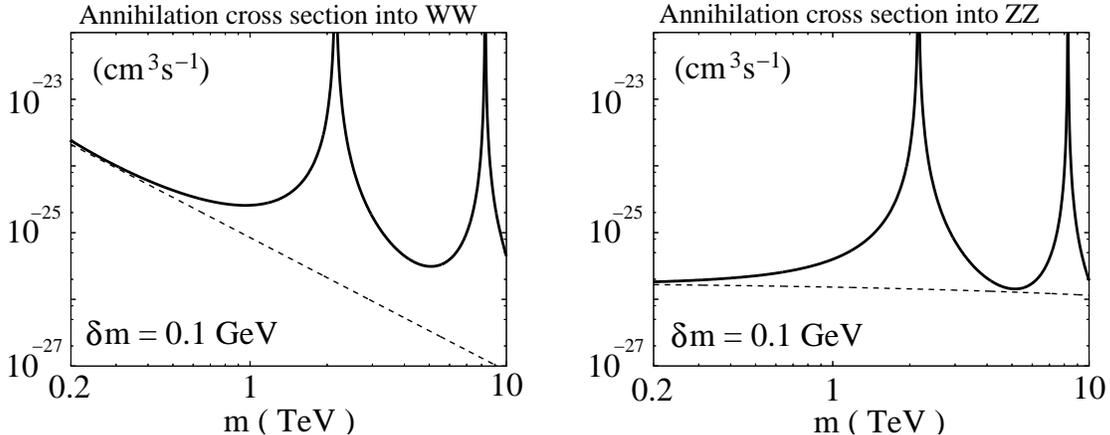}}
  \caption{ Cross sections, $\sigma v$, of the annihilation of the
    wino-like neutralinos into $W^+W^-$ (left figure) and $ZZ$ (right
    figure) in a non-relativistic limit. The mass difference between
    the wino-like neutralino and chargino is set to be 0.1~GeV. For comparison,
    the cross sections at the leading order in perturbation are shown
    as dashed lines.  The bound state resonances appear around 2~TeV
    and 8~TeV.  }
  \label{Fig:crosssection}
 \end{center}
\end{figure}

In Fig.~\ref{Fig:crosssection}, the annihilation cross sections into
$W^+W^-$ and $ZZ$ bosons are shown as functions of the wino mass.
These figures are plotted using fitting formulae for the wino-like
neutralino annihilation cross sections given in
Ref.~\cite{Hisano:2004ds}.  
When the mass difference between the wino-like neutralino and
chargino is much smaller than $~\alpha_2 m_W$, which is a typical
potential energy due to the electroweak interaction, the cross
sections are less sensitive to the value of the mass difference.  In
this paper, the mass difference is set to be 0.1~GeV for
definiteness.  For heavy wino-like neutralino, this mass difference
is dominated by the radiative correction, and it is $0.1-0.2$~GeV in
most of the parameters region \cite{Hisano:2004ds}.  This is because
the tree-level contribution to the mass difference is suppressed by
$({m_W}/{M_{\rm SUSY}})^4$ unless the wino mass is accidentally
finely tuned to the Higgsino mass. The mixing between the wino and
Higgsino components is also suppressed by $({m_W}/{M_{\rm SUSY}})$.
Thus, we ignore the mixing in the following.

As shown in the figure, the annihilation cross section into $ZZ$ is
also enhanced for $m\gsim 1$~TeV in addition to that into $W^+W^-$,
and it becomes comparable to that into $W^+W^-$. The cross sections
into $\gamma\gamma$ and $\gamma Z$ also have a behavior similar to
that into $ZZ$.  The annihilation channels into $ZZ$, $\gamma Z$ and
$\gamma\gamma$ come from one-loop diagrams in the perturbation, and
the cross sections are suppressed. However, the
transition between the neutralino pair state and the chargino pair
state is not suppressed due to the non-perturbative effect for $m\gsim
1$~TeV, so that the cross sections are enhanced.  When evaluating the
positron and antiproton fluxes from the wino-like neutralino
annihilation in the galactic halo, we need to include the contribution
of the annihilation into $Z$ bosons, in addition to that into $W$
bosons.

If the relic abundance of the wino-like neutralino in the universe is
explained by the thermal scenario, the mass consistent with the WMAP
observation is around 2~TeV \cite{Profumo:2004at}.  It is intriguing
that this value is coincident with the mass corresponding to the
resonant annihilation as shown in Fig.~\ref{Fig:crosssection}. On the
other hand, the wino-like neutralino DM is also produced by
non-thermal processes such as the moduli decay
\cite{Moroi:1999zb,Kohri:2005ru}. Furthermore, the late time entropy
production by, for example, the thermal inflation
\cite{Yamamoto:1985mb} may decrease the amount of the DM.  In these
cases, the mass of the wino-like neutralino consistent with the DM
observations may be deviated from 2~TeV.

In this paper, while the heavy wino-like neutralino with mass around
2~TeV is noticed, we discuss the positron and antiproton signatures
from the neutralino annihilation without peculiar masses specified for
completeness. Thus, we assume that the wino-like neutralino is
dominant constituent of the CDM in the present universe, and exists in
the halo of our galaxy with appropriate mass density in the following.

\section{Positron signature of wino-like neutralino dark matter}

In this section, we evaluate the positron flux from the wino-like
neutralino annihilation in the galactic halo. In the evaluation of the
signal flux in the vicinity of the solar system, we need to consider
the propagation of positrons through the galaxy, in addition to the
production rate of the positrons from the annihilation in the halo.
We discuss these in order, and show the sensitivities of the
upcoming experiments, such as PAMELA and AMS-02, to the positron
signal by comparing the expected background originated from the
secondary production of the cosmic rays. The HEAT anomaly is also
discussed.

\subsection{Production rate of positrons from dark matter annihilation}

The production rate of positrons from the neutralino DM annihilation in
the galactic halo is given as
\begin{equation}
 Q(E,\vec{r})
 =
 \frac{1}{2} n^2(\vec{r})
 \sum_{f}
 \langle \sigma v \rangle _f
 \left(\frac{d N_{e^+}}{d E}\right)_f~~,
\end{equation}
where $n$ is the number density of the neutralinos in the galactic
halo, $\langle \sigma v \rangle_{f}$ is the annihilation cross section
into the final state $f$.  The fragmentation function
$(d N_{e^+}/d E)_f$ represents the number of positrons with energy $E$,
which are produced from the final state $f$.  The coefficient $1/2$
comes from the pair annihilation of the identical particles.

As discussed in the previous section, the wino-like neutralinos
annihilate into $W$ and $Z$ bosons. Positrons are produced
through the leptonic and hadronic cascade decays of the weak gauge
bosons, for example, $W^+ \rightarrow e^+\nu$, $W^+ \rightarrow
\mu^+\nu \rightarrow e^+\nu \bar{\nu}\nu$ or $W^\pm \rightarrow {\rm
  hadrons} \rightarrow \pi^+ \rightarrow \mu^+ \rightarrow e^+$.
These cascade decay processes for producing positrons are encoded
into the fragmentation functions $(d N_{e^+}/d E)_f$ $(f=WW$ and $ZZ$).
We ignore the contribution from the annihilation into $Z\gamma$,
since the contribution is less than about $10\%$.  We evaluate the
fragmentation functions using the HERWIG Monte-Carlo code
\cite{Corcella:2000bw} and derive the fitting functions as follows,
\begin{eqnarray}
 \left(\frac{d N_{e^+}}{dx}\right)_{WW}
 =
 \exp\left[WW(\ln(x))\right]~,
 \qquad
\left( \frac{d N_{e^+}}{dx}\right)_{ZZ}
 =
 \exp\left[ZZ(\ln(x))\right]~,
\end{eqnarray}
where $x = E/m$ and the functions, $WW(x)$ and $ZZ(x)$, are given by
\begin{eqnarray}
 WW(x)
 &=&
 - 2.28838 - 0.605364x - 0.287614x^2 - 0.762714x^3
 \nonumber \\
 &&
 - 0.319561x^4 -0.0583274x^5 - 0.00503555x^6 - 0.00016691x^7~,
 \nonumber \\
 ZZ(x)
 &=&
 - 2.75588 - 0.45725x - 0.141373x^2 - 0.905392x^3
 \nonumber \\
 &&
 -0.444098x^4 - 0.0936451x^5 - 0.00942148x^6 - 0.000369777x^7 ~.
\end{eqnarray}

\begin{figure}[t]
 \begin{center}
  \includegraphics[width=7.4cm]{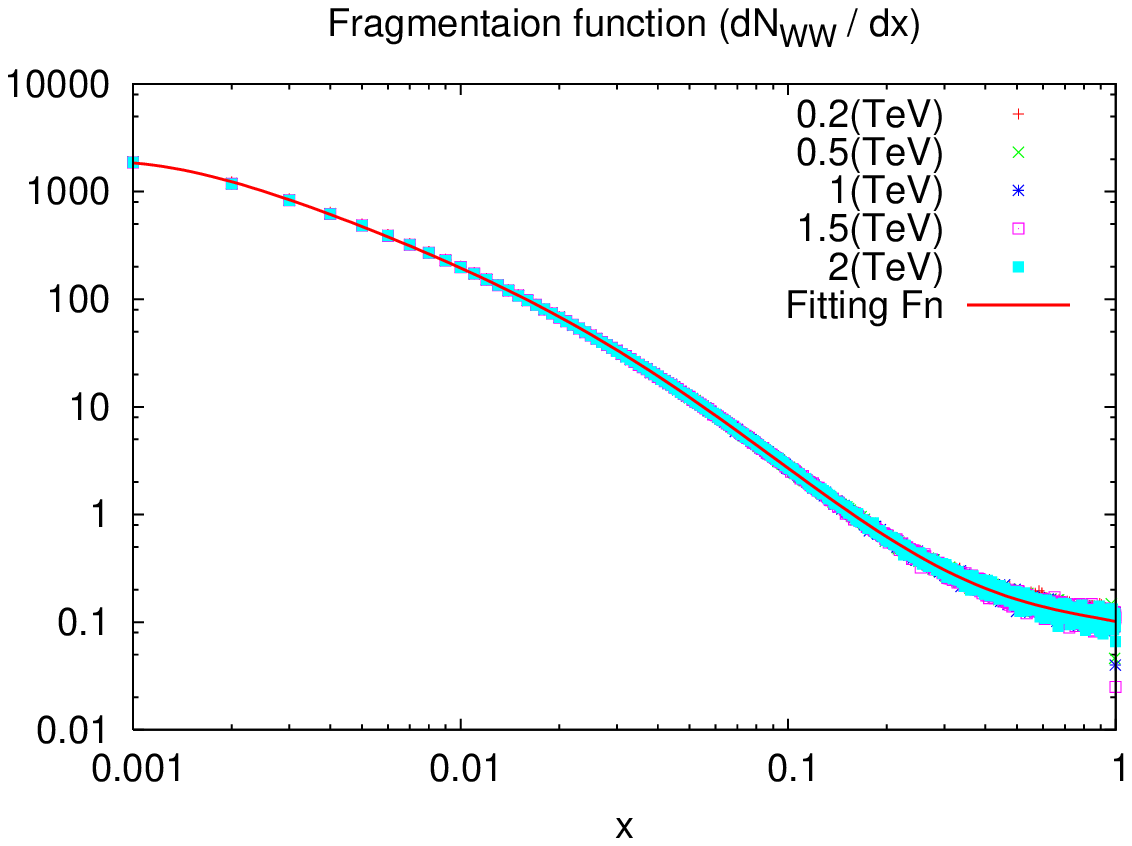}
  \includegraphics[width=7.4cm]{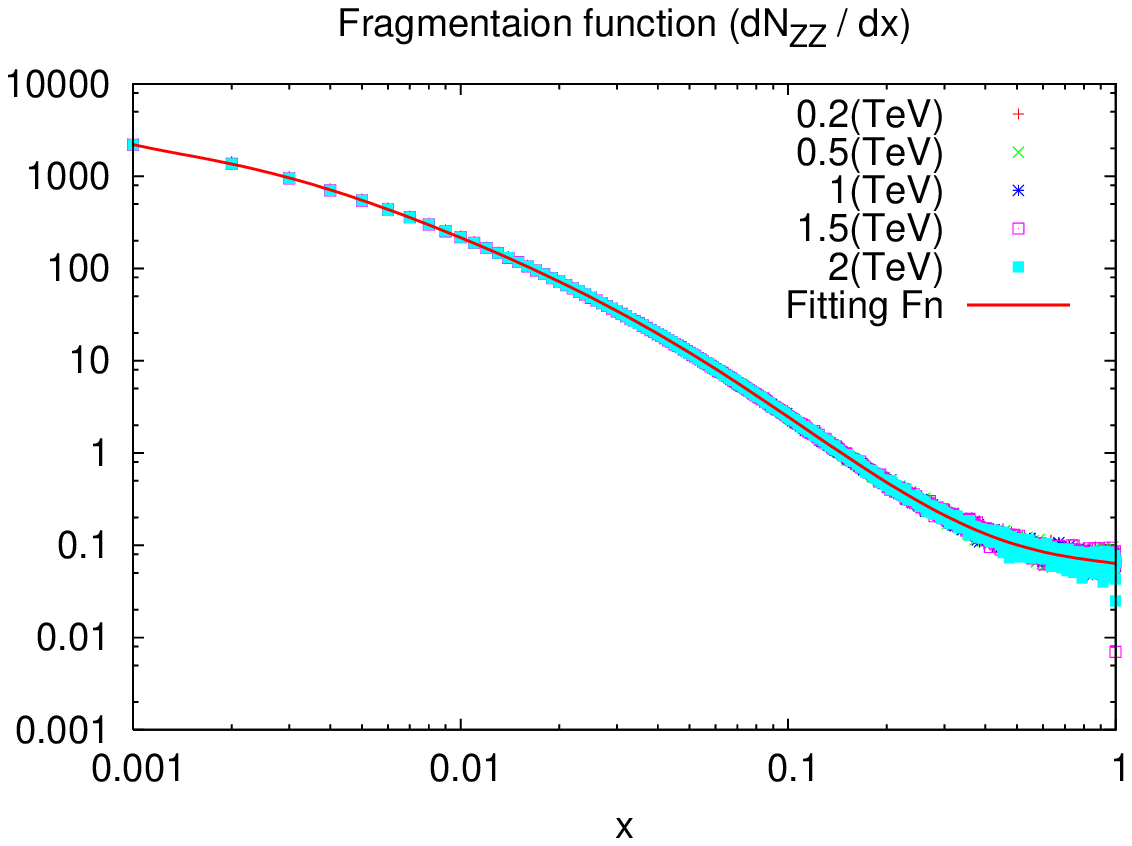}
  \caption{\small 
  Fitting functions of the fragmentation functions $(dN_{e^+}/dx)_{WW}$ and
  $(dN_{e^+}/dx)_{ZZ}$ (solid lines) and HERWIG Monte-Carlo results in cases of
  $m = 0.2, 0.5, 1, 1.5$, and 2~TeV.}
 \label{fitting}
 \end{center}
\end{figure}

In Fig.~\ref{fitting}, the fragmentation functions from the HERWIG
code and the fitting functions are depicted. The results of
Monte-Carlo simulations are shown for cases of $m = 0.2, 0.5, 1,
1.5$, and 2~TeV.  The fitting functions are shown as solid lines and
agree well with the simulation data with the range $m \gtrsim 300$~GeV
and $x \gtrsim 10^{-3}$.  It is found that the slopes of the
fragmentation functions are changed around $x \sim 0.2$.  The
positrons with lower energy $(x \lesssim 0.2)$ comes from the hadronic
cascade decay process \cite{Rudaz:1987ry}, while those with higher
energy $(x \gtrsim 0.2)$ are produced more directly from the leptonic
weak boson decays.

Next, we discuss the DM number density in the galactic
halo.  The number density is derived from the DM halo mass profile
 $\rho(\vec{r})$ through the equation $n(\vec{r}) =
\rho(\vec{r})/m$.  The halo mass profile is determined by observations
of the rotational velocity of the galaxy and the motions of the dwarf
galaxies with help of the $N$-bodies simulations, while several models for
the DM profile are proposed.  In this paper, we use the isothermal halo
model, which is given as 
\begin{equation}
 \rho(\vec{r})
 =
 0.43\frac{2.8^2 + 8.5^2}{2.8^2 + (r/1~{\rm kpc})^2}~~
 ({\rm GeV/cm^3})~,
\label{isothermal}
\end{equation}
where $r = |\vec{r}|$ is the distance from the galactic center,
0.43~GeV/cm$^3$ is the local halo density (the mass density in the
vicinity of the solar system), 2.8~kpc is the core radius of the
galaxy, and 8.5~kpc is the distance between the galactic center and the
solar system.

\subsection{Propagation of positrons in the galaxy}

Once positrons are produced by the DM annihilation, they
travel in the galaxy under the influence of the tangled magnetic
field.  Since the typical strength of the magnetic field is a micro
Gauss, the gyroradius of the positron is much less than the galactic
radius.  Thus, the propagation can be treated as a random walk, and
only some portion of the positrons can reach to the earth.

There are some models for the propagation. Among those, we use the
`diffusion model' in which the random walk is described by the diffusion
equation,
\begin{equation}
 \frac{\partial}{\partial t}f_{e^+}(E,\vec{r})
 =
 K(E)\nabla^2 f_{e^+}(E,\vec{r})
 +
 \frac{\partial}{\partial E}
 \left[b(E)f_{e^+}(E,\vec{r})\right]
 +
 Q(E,\vec{r})~,
 \label{eq:diffusion}
\end{equation}
where $f_{e^+}(E,\vec{r})$ is the number density of positrons per unit
energy, $E$ is the energy of positron, $K(E)$ is the diffusion
constant, $b(E)$ is the energy loss rate, and $Q(E,\vec{r})$ is the
source (positron injection) term discussed in the previous section.
The flux of positrons with high energy ($E \gg m_e$) in the vicinity
of the solar system is given from $f_{e^+}(E,\vec{r})$ as
\begin{equation}
 \Phi_{e^+}(E)
 =
 \frac{c}{4\pi}f_{e^+}(E,\vec{r}_\odot)~,
 \label{flux}
\end{equation}
where $c$ is the velocity of light and $\vec{r}_\odot$ represents
the coordinate of the solar system.

The diffusion constant $K(E)$ in Eq.~(\ref{eq:diffusion}) is obtained
by the simulation of cosmic rays, in which the diffusion model is
used. In particular, the Boron to Carbon ratio $B/C$ is an important
quantity for the simulation. By comparing the measurement of $B/C$ in
the cosmic rays and the result of the simulation, the diffusion
constant is evaluated. For the calculation of the positron flux,
we use the value in Refs.~\cite{Baltz:1998xv,Webber:1991},
\begin{equation}
 K(E)
 =
 3 \times 10^{27}
 \left[3^{0.6}+ (E / 1~{\rm GeV} )^{0.6}\right]~~
 ({\rm cm^2s^{-1}})~,
\end{equation}
where the form of $K(E)$ affects low energy positron flux,
while high energy one which we are interested in
is almost independent of the choice of this parameter.

The positrons lose their energies by the inverse Compton scattering with
cosmic microwave radiation (and infrared photons from stars) and the
synchrotron radiation with the magnetic field during the propagation
in the galaxy.  Therefore, the energy loss rate $b(E)$ is determined by
the photon density, the strength of the magnetic field and the Thomson
scattering cross section.  We use the value of $b(E)$ in
Refs.~\cite{Baltz:1998xv,Longair:1994wu},
\begin{equation}
 b(E)
 =  10^{-16} (E / 1~{\rm GeV})^2~~
 ({\rm  GeV s^{-1}})~.
\end{equation}

It is plausible that the positrons from the DM annihilation are in the
equilibrium in the present universe, and hence the number density
$f_{e^+}(E,\vec{r})$ is obtained by solving Eq.~(\ref{eq:diffusion}) with
the steady state condition $\partial f_{e^+}/\partial t = 0$.  Furthermore,
we impose the free escape boundary condition, namely the positron
density drops to zero on the surface of the diffusion zone.  The
positrons coming from the outside of the diffusion zone are
negligible, and the positrons produced inside the diffusion zone
contribute to the flux around the solar system, since they are trapped
due to the magnetic field \cite{Barrau:2001ev}.

It is usually assumed that the diffusion zone is a cylinder and that
its half-height and radius are $ L \sim (2 - 15)$~kpc and $ R =
20$~kpc, respectively. We fix $L=4$~kpc in the evaluation of the
positron flux.  However, high-energy positrons, which we interest,
only come from within a few kpc of the solar system as will discussed
later.  Hence, the positron flux is weakly dependent on the choice of
the parameters of the diffusion cylinder.  A detailed method for
solving the diffusion equation (\ref{eq:diffusion}) is presented in
Appendix~A.

Here we discuss the effect of the solar modulation on the positron
flux.  The flux given by Eq.~(\ref{flux}) is not exactly one to be
measured on the top of atmosphere.  The spectrum of the
interstellar flux in Eq.~(\ref{flux}) is modified due to interaction
with the solar wind and the magnetosphere.  However, the effect is not
so important when the energy of the positron is above 10~GeV.
Furthermore, the solar modulation effect is removed in the positron
fraction, that is a ratio of positron to the sum of positron and
electron fluxes, $e^+ /(e^+ +e^-)$.  Thus, we present our result
mainly in terms of the positron fraction.

\subsection{Background fluxes of positrons and electrons}

Positrons in the galaxy are injected by not only the DM annihilation
but also the scattering of cosmic-ray protons with the intersteller medium.
(See e.g. \cite{Moskalenko:1997gh}.)
The flux of these positrons is calculated by simulations,
in which the diffusion model is also used. The results agree with the
measurements of the low-energy positron flux in the cosmic rays
\cite{Moskalenko:1997gh}.

Since we can not distinguish the signal positrons,
which originate from the DM annihilation,
from those background positrons in measurements,
we need to know the background positron flux.
The background electron flux is also required for predicting the signals
in terms of the positron fraction.  In this paper, we use the fitting
functions of these background fluxes, which are obtained by the cosmic
ray simulations \cite{Baltz:1998xv},
\begin{eqnarray}
 \Phi^{\rm (prim)}_{e^- }(E)
 &=&
 \frac{0.16 E^{-1.1}}
      {1+11 E^{0.9}+3.2 E^{2.15}}~~
      ({\rm GeV^{-1}cm^{-2}s^{-1}sr^{-1}})~,
 \nonumber \\
 \nonumber \\
 \Phi^{\rm (sec)}_{e^{-}}(E)
 &=&
 \frac{0.70 E^{0.7}}
 {1+110 E^{1.5}+600 E^{2.9}+580 E^{4.2}}~~
 ({\rm GeV^{-1}cm^{-2}s^{-1}sr^{-1}})~,
 \nonumber \\
 \nonumber \\
 \Phi^{\rm (sec)}_{e^+}(E)
 &=&
 \frac{4.5 E^{0.7}}
 {1+650E^{2.3}+1500 E^{4.2}}~~
 ({\rm GeV^{-1}cm^{-2}s^{-1}sr^{-1}})~,
 \label{background}
\end{eqnarray}
where $E$ is in unit of GeV.  The first one, $\Phi^{\rm (prim)}_{
  e^-}$, is the flux of the primary electrons.  These electrons are
considered to be produced by the shock wave acceleration in
supernovae. On the other hand, the second and third ones, $\Phi^{\rm
  (sec)}_{ e^-}$ and $\Phi^{\rm (sec)}_{ e^+}$, are the secondary
electron and positron fluxes, respectively,
which are produced by the collisions of cosmic ray protons and helium nuclei
with hydrogen and helium of interstellar medium.

\subsection{Positron signature from dark matter annihilation}

In this section, we present the signature of the positrons from the
wino-like neutralino DM annihilation. The positron flux from heavy
 DM annihilation ($m\gtrsim$~1~TeV) is usually expected to be small. This
is because the source injection $Q$ scales as $\propto m^{-4}$ due to
the mass dependence of the cross section ($\propto m^{-2}$) and that
of the number density squared ($\propto m^{-2}$).  However, the mass
dependence of the cross section is very different from the ordinary
one when the DM is the wino-like neutralino as discussed in the previous
section. Furthermore, the cross section is enhanced by several orders
of magnitude when the neutralino has the mass around 2~TeV. Thus, the
positron flux is expected to be large in this case.

\begin{figure}[t]
 \begin{center}
  \includegraphics[width=10cm]{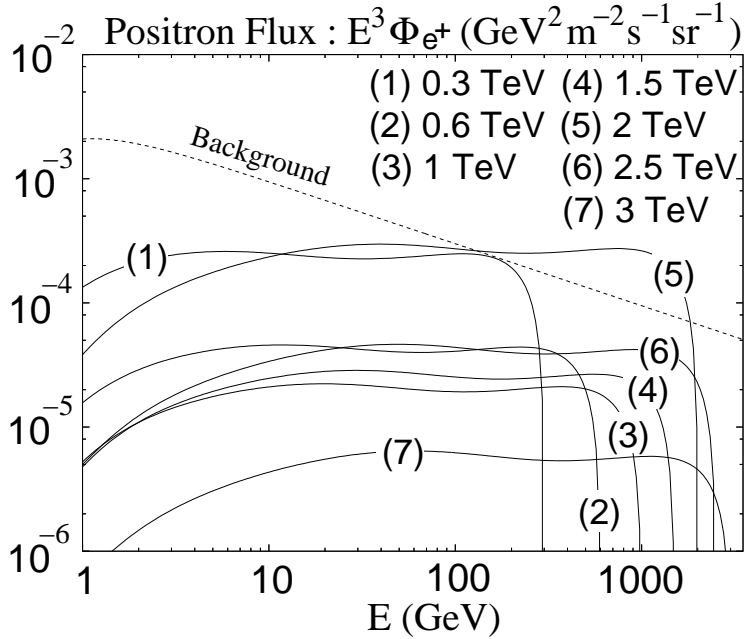}
  \caption{\small (Interstellar) positron flux from the wino-like
    neutralino annihilation.  The signal fluxes for the wino mass
    $m=0.3$, 0.6, 1, 1.5, 2, 2.5, and 3~TeV are shown as solid lines.
    The expected background flux of positrons from the cosmic ray
    simulation is also shown as a dotted line.  }
 \label{positron flux}
 \end{center}
\end{figure}

First, we show the positron flux from the wino-like neutralino
annihilation in Fig.~\ref{positron flux}. In this figure, the signal
flux is shown as solid lines. The wino mass is taken to be $m=0.3$,
0.6, 1, 1.5, 2, 2.5, and 3~TeV. For comparison, the expected
background flux of positrons from the cosmic ray simulation is
also shown as a dotted line.  The effect of the solar modulation is
not included, and thus the spectrums below 10~GeV have uncertainties.
However, since the high-energy positron spectrum is important for the
discrimination of the signal from the background as indicated in
Fig.~\ref{positron flux}, the uncertainties from the solar modulation
are not serious.

When the wino mass is around 300~GeV, the signal flux is comparable to
the background flux in the energy range 100~GeV$\lsim E\lsim$ 300~GeV.
Furthermore, the signal flux for the mass around 2~TeV also exceeds
the background one in the energy range $E \gtrsim 100$~GeV.  The
latter comes from the resonant DM annihilation. It is also noticed
that a bump appears in each signal spectrum at around $m/2$.  The
positrons with energy above the bump come from the direct decay of 
weak gauge bosons, while those with energy below the bump are produced
mainly by the hadronic cascade decay of the gauge bosons.

\begin{figure}[t]
 \begin{center}
   \includegraphics[width=7.2cm]{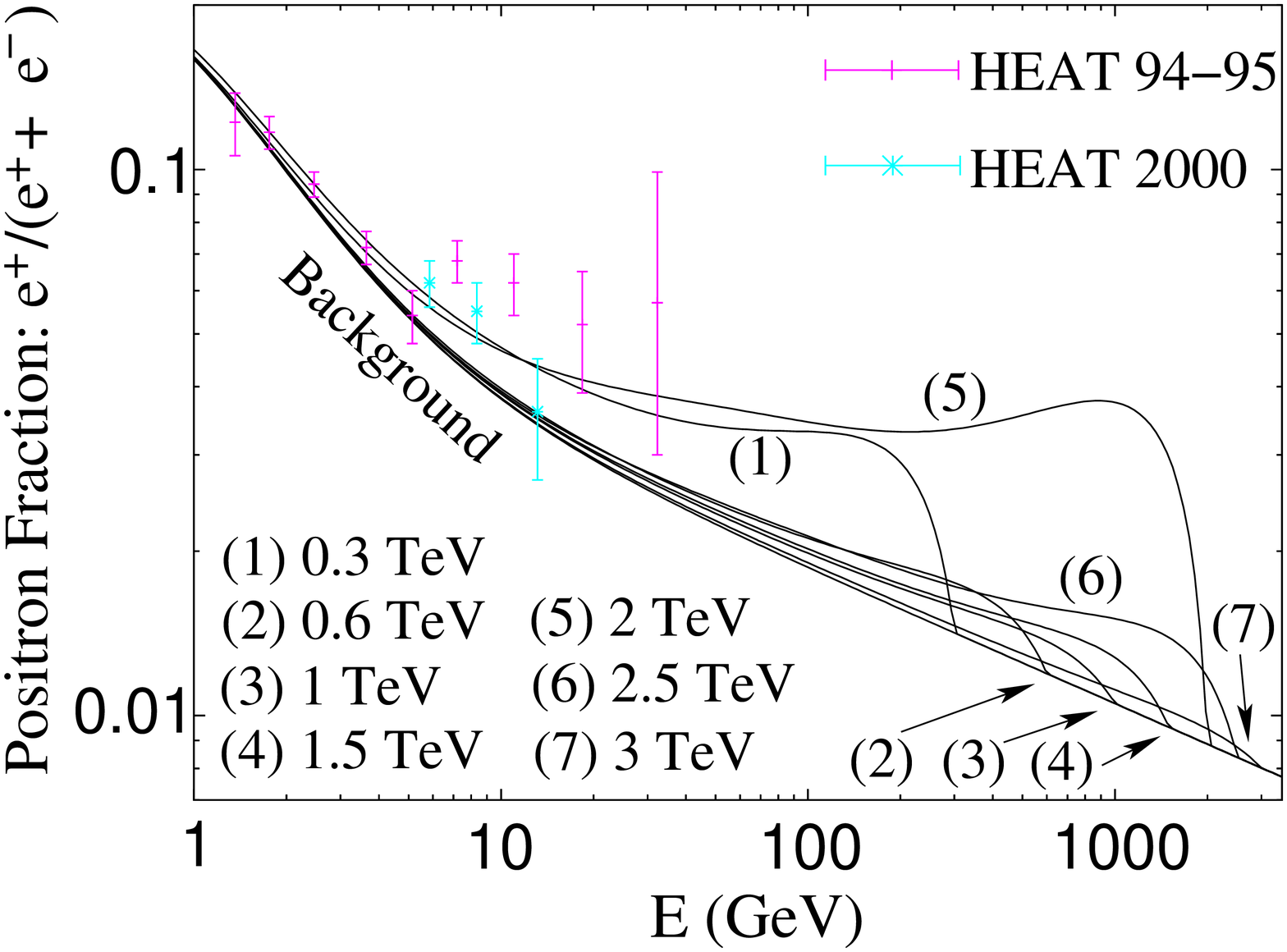}
  ~
  \includegraphics[width=7.0cm]{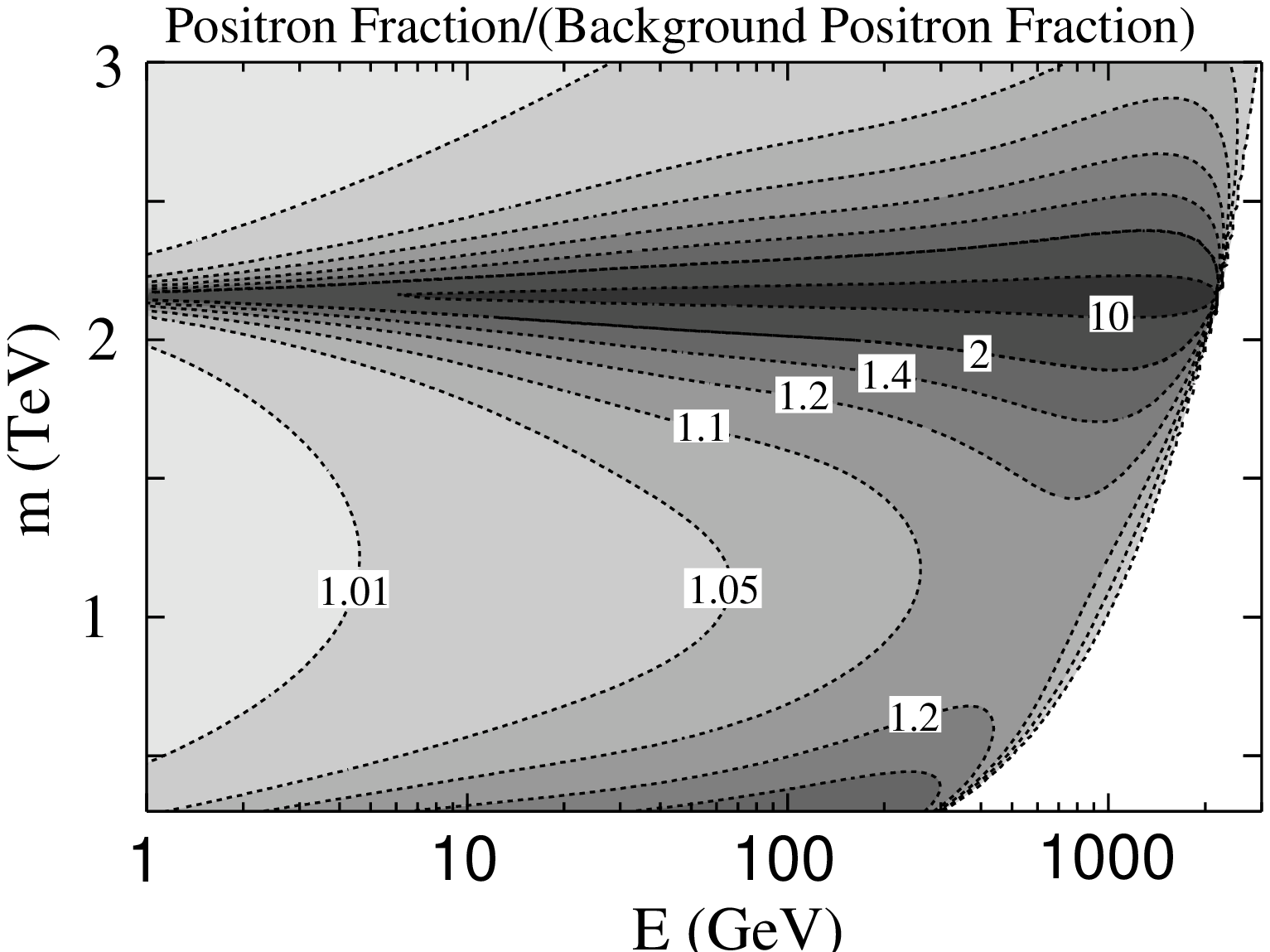}
  \caption{\small (Left figure) Positron fraction, $e^+ /(e^+ +e^-)$,
    as a function of positron energy $E$ in the wino-like neutralino DM.
    For comparison, the expected background positron fraction,
	the positron data HEAT 94-95 and HEAT 2000 are also shown in this figure.
	(Right figure) Contour plot of the ratio
    between the positron fractions including positrons from the DM
    annihilation and without it (that is the background
    positron fraction) in a $(E,m)$ plane.  }
  \label{positron fraction}
 \end{center}
\end{figure}

Next, we consider the positron fraction calculated from the positron
flux in the Fig.~\ref{positron flux} and the expected background
ones in Eqs.~(\ref{background}).  The result is shown in
Fig.~\ref{positron fraction}. In the left figure, the positron
fraction is depicted as a function of positron energy for several
wino masses.  The choice of the mass is the same as that in the
Fig.~\ref{positron flux}.
The expected background positron fraction,
the positron data HEAT 94-95 \cite{Barwick:1997ig}
and HEAT 2000 \cite{Beatty:2004cy} are also shown in this figure.
In the right figure of
Fig.~\ref{positron fraction}, the ratio of the fraction including
positrons from the DM annihilation to the background one
is depicted as a contour plot in a $(E,m)$ plane.
From these figures, it is clear that the signature becomes more significant
for high-energy positrons. In particular, there is a large difference
between the expected signal and the background when the wino mass is a
few hundred GeV or around 2~TeV.

Here, we address the HEAT experiment \cite{Barwick:1997ig,Beatty:2004cy},
which reported the positron excess from the expected background.
The spectrum of the observed fraction is almost flat around 0.06 in a
energy range 4~GeV $\lesssim E \lesssim$ 20~GeV.  The positron
fractions for both $m=300$~GeV and $2$~TeV in the figure are
consistent with it within the experimental error.

In addition, the effect of the inhomogeneity in the local DM
distribution on the positron flux is recently discussed, whose
existence is supported by the $N$-bodies simulations. In these
arguments, the positron flux from the DM annihilation is enhanced if
there are clumps of the DM in the vicinity of the solar system. The
effect is parametrized as a boost factor ($BF$)
 \cite{boostfactor}, which is defined by a
ratio of the signal fluxes with inhomogeneity and without
inhomogeneity. The boost factor may reach $\sim$ 5 when the
inhomogeneity exists, while the factor is equal to one if the DM is
distributed homogeneously. Thus, the wino-like neutralino with the
mass $\sim$ 300~GeV or 2~TeV can explain the HEAT result quite naturally.
It is amazing that the wino-like neutralino with 2~TeV naturally
accounts for not only the DM abundance thermally but also the HEAT
anomaly.

\begin{figure}[t]
 \begin{center}
   \includegraphics[width=10cm]{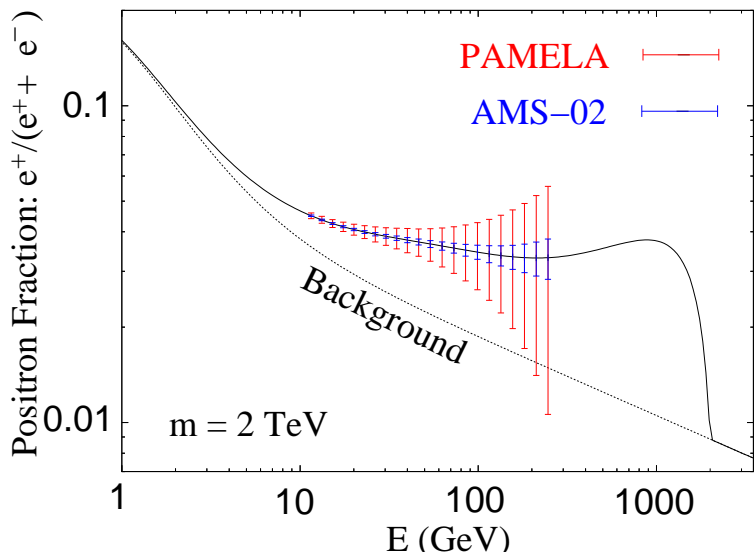}
  \caption{\small Sensitivities of the upcoming experiments.
	The positron fraction, $e^+ /(e^+ +e^-)$, for $m = 2$~TeV and
	that of the background are shown as solid and dotted lines, respectively.
	The error bars in the figure correspond to the statistical errors
	projected for the PAMELA and AMS-02 experiments
	after three years of observations.}
  \label{future-exp}
 \end{center}
\end{figure}

Next, we discuss the potential of the upcoming PAMELA \cite{Boezio:2004jx}
and AMS-02 \cite{Barao:2004ik} experiments,
which have good sensitivities in a broad region 
of positron energy 10~GeV $\lesssim E \lesssim$ 270~GeV,
might detect the signal from the wino-like neutralino DM annihilation.
We estimate the sensitivities of those experiments
following the method in Ref.~\cite{Hooper:2004bq}.
In Fig.~\ref{future-exp}, we show the sensitivities of the PAMELA and AMS-02.
The positron fraction, $e^+ /(e^+ +e^-)$, for $m = 2$~TeV and
that of the background are shown as solid and dotted lines, respectively.
The error bars in the figure correspond to the statistical errors
projected for the PAMELA and AMS-02 experiments
after three years of observations.
As shown in this figure, positrons with energy of some tens of GeV
will be clearly discriminated from the background.

Finally, we discuss other uncertainties of the signal flux.  First, in
the case of the positron propagation with high energy, we do not have
to worry about uncertainties from the thickness of the tangled
magnetic field ($L$).  This is because high-energy positrons we observe are
produced within a few kpc around the solar system.  Positrons far
from the earth lose their energies during the propagation, and
consequently they contribute to the low-energy part of the flux. The
distance in which positrons travel without significant energy loss is
typically
\begin{eqnarray}
r&\simeq&\sqrt{\frac{K(E) E}{b(E)}}
= 1.7\times\left(E/{\rm 100~GeV}\right)^{-0.27}~({\rm kpc})~.
\end{eqnarray}
Thus, the positron flux at high energy does not suffer from the
uncertainties of the thickness $L$ (because $L \gtrsim$ a few kpc).

Second is the DM distribution in the halo.  We have assumed the
isothermal halo in Eq.~(\ref{isothermal}) in the above. Various DM
halo models are proposed from the $N$-bodies simulations, however, the
high-energy positron flux from the DM annihilation is considered to be
almost independent of the choice of the halo model. The main
difference among the halo models appears in the galactic center.
However, the high-energy positrons produced around the galactic center
can not reach to the earth, and the positron flux has little ambiguity
from it around the solar system.

\section{Antiproton signature from wino-like dark matter annihilation}

The antiproton flux from the wino-like neutralino DM annihilation is
discussed in this section. The method for calculation of the flux is
essentially the same as that of the positron flux. First, the
antiproton injection in the galactic halo (source term) and the
propagation of the antiprotons are discussed, and the antiproton
flux from the wino-like neutralino annihilation in the galactic
halo is evaluated. The antiproton background originated from the
cosmic rays is also discussed.

\subsection{Production rate of antiprotons from dark matter
  annihilation}

Antiprotons from the wino-like neutralino annihilation are also
produced through the cascade decay of weak gauge bosons. The
difference between the antiproton and the positron production rates
(source terms) appears only in the fragmentation functions, and then
the antiproton production rate is given as
\begin{equation}
 Q(T,\vec{r})
 =
 \frac{1}{2}
 n^2(\vec{r})
 \sum_{f}
 \langle \sigma v \rangle _f
 \left(\frac{dN_{\bar p}}{d T}\right)_f~.
 \label{Q term}
\end{equation}
where $T(\equiv E - m)$ is the kinetic energy of antiproton
and $(dN_{\bar p}/dT)_f$ is the fragmentation function.  

\begin{table}
 \begin{center}
  {\scriptsize
  \begin{tabular}{|l|c c c c|}
   \hline
         & $j=1$   & $j=2$   & $j=3$                  & $j=4$\\
   \hline
   $i = 1$ & 306.0 & 0.28  & 7.2$\times 10^{-4}$  & 2.25 \\
   $i = 2$ & 2.32  & 0.05  & 0                    & 0 \\
   $i = 3$ & -8.5  & -0.31 & 0                    & 0 \\
   $i = 4$ & -0.39 & -0.17 & $-2.0\times 10^{-2}$ & 0.23 \\
   \hline
  \end{tabular}
  \qquad\qquad
  \begin{tabular}{|l|c c c c|}
   \hline
         & $j=1$   & $j=2$   & $j=3$                  & $j = 4$ \\
   \hline
   $i = 1$ & 480.0 & 0.26   & 9.6$\times 10^{-4}$  & 2.27 \\
   $i = 2$ & 2.17  & 0.05   & 0                    & 0 \\
   $i = 3$ & -8.5  & -0.31  & 0                    & 0 \\
   $i = 4$ & -0.33 & -0.075 & $-1.5\times 10^{-4}$ & 0.71 \\
   \hline
  \end{tabular}
  }
  \caption{\small Coefficients in Eq.~(\ref{pim}), $a_{ij},$ for
    $W^{+}W^{-}$ process (left panel) and $ZZ$ one (right panel).}
 \label{aij for antip}
 \end{center}
\end{table}

As in the case of the fragmentation functions for positrons, the
functions for antiprotons are calculated from the Monte-Carlo
simulation.  In this paper, we use the simple parametrization in
Ref.~\cite{Bergstrom:1999jc} which fits the result of the PYTHIA
Monte-Carlo code \cite{Sjostrand:1993yb},
\begin{equation}
 \left(\frac{dN_{\bar p}}{dx}\right)_f
 =
 \left(p_1 x^{p_3} + p_2|\log_{10}x|^{p_4}\right)^{-1}~,
\end{equation}
where $x = T/m$. The parameters $p_i$ in the above equation depend on
the neutralino mass in addition to the annihilation channels, and they
are given as
\begin{equation}
 p_i(m)
 =
 \left(a_{i1}m^{a_{i2}} + a_{i3}m^{a_{i4}}\right)^{-1}~.
\label{pim}
\end{equation}
The values of the coefficients, $a_{ij}$, are listed in Table~\ref{aij
  for antip} for the $W^{+}W^{-}$ process (left panel) and the $ZZ$
one (right panel).  The parametrization for the fragmentation
functions is valid for the neutralino mass in the range (50 --
5000)~GeV. We dropped quark processes such as $t \bar t$ and $b \bar b$
since the annihilation cross sections are very small
due to heavy squark masses and helicity suppression.
  
\subsection{Propagation of antiprotons in the galaxy}

In order to treat the propagation of antiprotons, we use the diffusion
model as in the case of positrons. The diffusion equation describing
the propagation is written as
\begin{eqnarray}
 K_{p}(T)\nabla^2 f_{\bar p}(T,\vec{r})
 - \frac{\partial}{\partial z} \left( V_C(z)f_{\bar p}(T,\vec{r}) \right)
- 2 h \delta(z)\Gamma_{\rm ann} f_{\bar p}(T,\vec{r})
&& \nonumber \\ 
 + Q(T,\vec{r}) + Q^{\rm tert} (T,\vec{r})
 &=& 0~,
 \label{eq:diff}
\end{eqnarray}
where $f_{\bar p}(T,\vec{r})$ is the number density of antiprotons
per unit energy. The steady state condition ($\partial f_{\bar p}
/ \partial t = 0$) is assumed as discussed in the positron case.  For
the evaluation of the equation, we use the cylinder coordinate.  The
interaction of antiproton with matter is confined on the galactic
plane, which is expressed as the infinitely thin disk with radius
$R=20$~kpc at the $z = 0$.  The diffusive halo is the cylinder with
radius $R=20$~kpc and the half-height $L$.  The boundary condition for
solving the equation is taken to be the same as the positron case.

The diffusion equation (\ref{eq:diff}) is essentially the same as that
in Eq.~(\ref{eq:diffusion}).  However, there are some differences, for
example, the energy-loss term does not appear in
Eq.~(\ref{eq:diff}). This is because protons are much heavier than
electrons, so that we can neglect the energy-loss due to the
scattering with background photons.  The other differences are the
term related to the convective wind (second term), the interaction
term with matter in the galactic plane (third term) and the tertiary
antiproton term (last term).  These three terms are not so important
when we consider the antiproton flux with high energy ($T \gtrsim $ a
few GeV).  We include these terms in the diffusion equation for
completeness.

The diffusion coefficient $K_{p}$ is determined by the Boron to
Carbon ratio $B/C$ in the cosmic rays, which is the same as the
positron case.  For the calculation of the antiproton flux, we
parametrize the diffusion constant as
Refs.~\cite{Maurin:2002ua,Maurin:2002hw},
\begin{equation}
 K_{p}(\mathcal{R})
 = K_0 \beta \mathcal{R}^{\delta}~,
\end{equation}
where the diffusion coefficient is assumed to be constant within the
diffusion zone.  The variable $\mathcal{R}$ is called the rigidity,
which is defined by the momentum of the particle per unit charge
$\mathcal{R}\equiv p/Z$.  For the
values of $\delta$ and $K_0$, we use the parameter sets in
Table~\ref{table:galaxypara}.  These values are favored from the $B/C$
analysis \cite{Donato:2003xg}.

\begin{table}
 \begin{center}
  {\small
  \begin{tabular}{c|c c c c }
   \hline
   case  & $\delta$ & $K_0$(kpc$^2$/Myr) & $L$ (kpc) & $V_c$(km/s)\\
   \hline
   max & 0.46  & 0.0765 & 15 & 5 \\
   med & 0.70  & 0.0112 & 4  & 12 \\
   min & 0.85  & 0.0016 & 1  & 13.5 \\
    \hline
  \end{tabular}
  }
  \caption{\small Astrophysical parameters compatible with the $B/C$ analysis
\cite{Donato:2003xg}.
Three cases give the maximal, median and minimal signal antiproton fluxes.
} \label{table:galaxypara}
 \end{center}
\end{table}

The second term in Eq.~(\ref{eq:diff}), $\partial (V_C(z)f_{\bar p})
/ \partial z$, is not included in the equation for the positron flux.
This term is related to the convective wind, which represents the
movement of medium responsible for the diffusion.  The direction of
the wind is assumed to be perpendicular to the disc plane, and the
velocity $V_C(z)$ is constant throughout the diffusive volume,
\begin{equation}
 V_C(z)
 = \left(2\theta (z) - 1\right)V_c~,
\end{equation}
where the value for $V_c$ is
given in Table~\ref{table:galaxypara}.

Next one is the third term in Eq.~(\ref{eq:diff}), $-2 h \delta(z)
\Gamma_{\rm ann} f_{\bar p}$, which represents the annihilation
between antiproton and interstellar proton in the galactic plane.
The parameter $h$ in the term is the half-height of disk and set to be
100~pc $(\ll L)$, while $\Gamma_{\rm ann}$ is the annihilation rate
between antiproton and proton,
\begin{equation}
 \Gamma_{\rm ann}
 = ( n_{\rm H} + 4^{2/3} n_{\rm He} ) \sigma^{\rm ann}_{\bar{p}p}v_{\bar{p}}~,
\end{equation}
where $v_{\bar{p}}$ is the velocity of antiproton,
$n_{\rm H} $ denotes the hydrogen number density ($\sim$ 1~cm$^{-3}$),
and $n_{\rm He}$ is the helium number density
which we assume to be 7\% of $n_{\rm H}$ \cite{Garcia:1987}. 
The factor $4^{2/3}$ arises from a geometrical approximation \cite{Donato:2001ms}.
The annihilation cross section between antiproton and proton,
$\sigma^{\rm ann}_{\bar{p}p}$, is given by \cite{Tan:1983de,Protheroe:1981gj}
\begin{eqnarray}
\sigma^{\rm ann}_{\bar{p}p}(T)
 = \left\{
  \begin{array}{ll}
  661 \left( 1 + 0.0115 ~ T^{-0.774} - 0.948 ~ T^{0.0151} \right) ~ {\rm mb} ~ ,
 & T < 15.5 {\rm~GeV}~, \\
  36 ~ T^{-0.5} ~ {\rm mb} ~ ,
 & T \geq 15.5~{\rm GeV}~, \\
   \end{array} \right. 
   \label{eq:ann}
\end{eqnarray}
where $T$ is in unit of GeV.  This interaction dominates over
inelastic interactions at low energy.  Hence, the flux of
antiprotons with low energy is decreased by the annihilation.

For higher energy antiprotons ($T \gtrsim 10$~GeV), the inelastic
interaction is not dominated by annihilation, however, the
non-annihilating scattering is important.  The interaction lowers
energies of antiprotons, $T'$ to $T (< T')$.  These antiprotons are
called tertiary antiprotons.  We include this effect in $Q^{\rm tert}
(T,\vec{r})$, which is given by
\begin{eqnarray}
Q^{\rm tert} (T,\vec{r}) &=& ( n_{\rm H} + 4^{2/3} n_{\rm He} )
\nonumber \\
&& \times \left[ \int_{T}^m \frac{\sigma_{\bar p p}^{\rm non-ann} (T') }{T'}
 v'_{\bar p} f_{\bar p} (T',\vec{r}) d T'
- \sigma_{\bar p p}^{\rm non-ann}(T)
 v_{\bar p} f_{\bar p} (T,\vec{r}) \right] .
 \label{eq:tertiary}
\end{eqnarray}
The first term in the bracket is the contribution to the antiproton
flux with energy $T$ from the inelastic scattering of antiprotons with
energy larger than $T$, while the second term compensates it so that
the total antiproton number is not changed in this process. Here, $
\sigma_{\bar p p}^{\rm non-ann}(T) $ is given as the difference
between the total inelastic cross section $\sigma_{\bar p p}^{\rm
  inel}$ and the annihilation cross section $\sigma_{\bar p p}^{\rm
  ann}$.  The total inelastic cross section is given in
Ref.~\cite{Tan:1983de} as
\begin{equation}
 \sigma^{\rm inel}_{\bar p p}(T)
 = 24.7
 \left( 1 + 0.584 T^{-0.115}
  + 0.856 T^{-0.566} \right)~~({\rm mb})~,
  \label{eq:inelastic}
\end{equation}
where $T$ is in unit of GeV.

The number density of antiprotons $f_{\bar p}$ is obtained by
solving the diffusion equation (\ref{eq:diff}).  We can solve this
equation full-analytically \cite{Barrau:2001ev,Maurin:2002ua}.  The
detailed expression of the solution is presented in Appendix B.  After
solving the equation for the number density, the interstellar flux of
antiprotons from the DM annihilation in the vicinity of the solar
system is obtained as
\begin{equation}
 \Phi_{\rm IS}
 =  \frac{v_{\bar{p}}}{4\pi}f_{\bar p}(T,\vec{r}_\odot).
\end{equation}

Here, we discuss the effect of the solar modulation on the antiproton
flux.  This is important for antiprotons with low kinetic energies
($\lesssim$ 3~GeV). Using the force field approximation,
the flux of antiprotons on the top of atmosphere $\Phi_{\rm TOA}$ is
obtained from the interstellar flux $\Phi_{\rm IS}$ as
\begin{eqnarray}
 \frac{\Phi_{\rm TOA}(T^{\rm TOA})}
      {\Phi_{\rm IS}(T^{\rm IS})}
 =  \left(\frac{p^{\rm TOA}}{p^{\rm IS}}\right)^2~,
 \qquad T^{\rm TOA}
 = T^{\rm IS} - |Z| \phi~,
\end{eqnarray}
where $p^{\rm TOA}(T^{\rm TOA})$ and $p^{\rm IS}(T^{\rm IS})$ are
momentums (kinetic energies) of antiproton on the top of the
atmosphere and in the interstellar, respectively.  The solar modulation
parameter $\phi$ varies according to the 11 years solar cycle. This
parameter takes a value from about 500~MV at the minimum solar
activity to 1.3~GV at the maximum solar activity.  Larger $\phi$
lowers the antiprotons flux on the top of atmosphere flux at low
energy.

\subsection{Background flux of antiprotons}

In this section, we discuss the antiproton background flux. The
antiprotons are produced as the secondary products of cosmic rays by
the nuclear reaction with the interstellar gas in the galactic
disk. The main contribution to the antiproton flux comes from the
collision between the cosmic ray protons and the interstellar hydrogen
gas. Again, the production phenomena are described by the diffusion
equation. We solve the diffusion equation and calculate the background
flux. Since the concrete formalism for obtaining the flux is very
complex, we mention only the strategy for the calculation here.  The
antiproton background is also discussed in
Refs.~\cite{Bergstrom:1999jc,Donato:2001ms,Bottino:1998tw}.

While the interstellar primary proton flux is required to evaluate the
background antiproton flux, it is impossible to measure it directly.
However, it is obtained by solving the diffusion equation under an
assumption of the source function.  The primary protons are believed
to be produced by supernovae.  Hence, the proton source term with a
few undetermined parameters is assumed, and the interstellar proton
flux is obtained by solving the diffusion equation with this source
term.  In this case, the parameters in the source term are fixed by
comparing the evaluated flux with the observed cosmic rays on the
earth in the measurements such as BESS \cite{Sanuki:2000wh} and AMS
\cite{Alcaraz:2000ks}.  The fitting function for the primary proton
flux derived as above is given in Ref.~\cite{Donato:2001ms}. We use it in
our evaluation of the background antiproton flux.

Next, the antiproton flux is evaluated from the primary proton flux
by solving the corresponding diffusion equation. The equation is the
same as that in Eq.~(\ref{eq:diff}) except for the source term. Since
the antiprotons are produced by the nuclear reaction between the
cosmic rays and interstellar gas, the source term is given by the
proton flux and the cross sections for the reactions.

The antiprotons are dominantly produced by the process $p + {\rm H}
\rightarrow \bar{p} + {\rm X}$.  In the rest frame of the hydrogen
atom, the kinetic energy threshold for the incident proton to produce
secondary antiprotons is $6m_p$.  Furthermore, the number density of
the incident proton decreases as energy increases. As a result, the
spectrum of antiprotons from this process has a peak at a few GeV.  In
addition to this process, we include the inelastic collision between
proton and helium, $p + {\rm He} \rightarrow \bar{p} + {\rm X}$, for
generating the secondary antiprotons.  The process with the helium
contributes to the antiproton flux sub-dominantly in the most energy
range.  However, the antiprotons from the process are a dominant
component at low energy with the tertiary antiprotons ($T \lesssim$
0.1~GeV). Thus, the source term for the secondary antiproton turns
out to be
\begin{equation}
 Q(T)
 =
 2\int_{T_{\rm th}}^\infty dT'
 4\pi\Phi_p^{\rm (prim)}(T')
 \left[
  n_{\rm H}\frac{d\sigma_{p{\rm H}\rightarrow\bar{p}X}}{dT'}(T',T)
  +
  n_{\rm He}\frac{d\sigma_{p{\rm He}\rightarrow\bar{p}X}}{dT'}(T',T)
 \right]~.
\end{equation}
The factor 2 comes from the fact that the antiprotons are produced
from the antineutron decay in addition to the direct production of
antiprotons. The threshold energy $T_{\rm th}$ is $6m_{p}$
and $\Phi_{p}^{\rm (prim)}$ is the proton flux.
The differential cross section ${d \sigma} (T',T) /
{dT'}$ is for the production of an antiproton with energy $T$ from an
incident proton of energy $T'$. The cross sections are given in
Ref.~\cite{Duperray:2003bd}.

Within uncertainties of the observations, the obtained flux for the
antiproton background is consistent with the results by BESS
\cite{Orito:1999re}, AMS \cite{Aguilar:2002ad} and CAPRICE
\cite{Boezio:2001ac}, which observe the low-energy antiprotons
((0.2-50)~GeV).  We use the background flux for estimating the
antiproton signature from the DM annihilation.

\subsection{ Antiproton signature from dark matter annihilation}

Now we are in a position to discuss the antiproton signature from the
wino-like neutralino annihilation in the galactic halo. We calculate
the antiproton flux from the neutralino annihilation by solving
Eq.~(\ref{eq:diff}), and compare the result with the background flux
discussed in the previous section.

\begin{figure}
 \begin{center}
  \includegraphics[width=9.1cm]{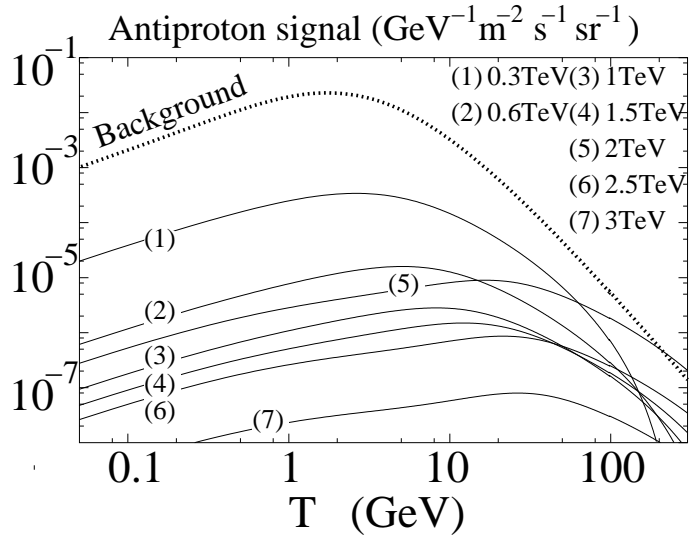}
  \caption{Antiproton flux from the wino-like neutralino annihilation
    on the top of atmosphere as a function of antiproton kinetic
    energy. The wino mass is taken to be $m=0.3$, 0.6, 1, 1.5, 2, 2.5,
    and 3~TeV. The solar modulation parameter $\phi$ is set to be
    500~MV.  For comparison, the background flux is depicted as a
    dotted line.}
 \label{pbar signals}
 \end{center}
\end{figure}

In Fig.~\ref{pbar signals}, the flux of the antiproton signal on the
top of the atmosphere is depicted for various wino masses as a
function of the antiproton kinetic energy.  In this figure, we use the
astrophysical parameters of the median set in
Table~\ref{table:galaxypara}, which gives the minimal $\chi^2$ for
the $B/C$ analysis \cite{Donato:2003xg}.  The parameter of solar
modulation $\phi$ is set to be 500~MV, which corresponds to almost
minimum solar activity.  For comparison, the background flux is also
shown as a dotted line.  As shown in this figure, it is implausible to
exceed the background for almost all region of the wino mass.  The
exception is only the case of the wino mass around 2~TeV. In this
case, the wino-like neutralinos annihilate resonantly as discussed in
section~\ref{sec:resonance}, and the signal flux is almost comparable
to the background flux at high energy ($T \gtrsim 100$~GeV).

\begin{figure}[t]
 \begin{center}
  \includegraphics[width=9.5cm]{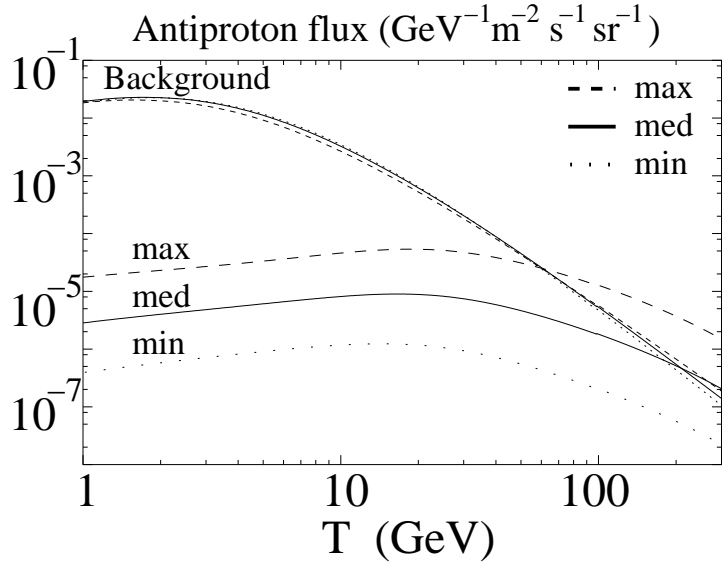}
  \caption{Antiproton flux from the wino-like neutralino annihilation
    and the background flux on the top of atmosphere for three
    astrophysical parameter sets in Table~\ref{table:galaxypara}.
    Here, we take the wino mass 2~TeV. The solar modulation parameter
    $\phi$ is set to be 500~MV as in the previous figure.}
 \label{pbar flux}
 \end{center}
\end{figure}

Let us discuss the uncertainties in the prediction of the signal
antiproton flux.  In Fig.~\ref{pbar flux} we show the signal
antiproton and the background fluxes on the top of atmosphere for
three astrophysical parameter sets in Table~\ref{table:galaxypara}, in
order to see the dependence on choice of astrophysical parameter sets.
Here, we take the wino mass 2~TeV and the solar modulation parameter
is 500~MV.

The uncertainties in the astrophysical parameters lead to an
uncertainty of a factor $O(100)$ for the prediction of the signal
antiproton flux. The flux from the neutralino annihilation depends
strongly on the astrophysical parameters, especially, the value of the
thickness of the diffusion zone $L$.  A larger $L$ means more
injection of signal antiprotons and leads to larger signal flux.  On
the other hand, the background flux has no strong sensitivity to $L$,
as shown in the figure. As a result, we can observe the signal
antiprotons only when $L$ is large enough.  The situation is very
different from the positron signal, in which only the DM within a few
kpc contributes to the flux due to the rapid energy loss of
high-energy electrons.

Other uncertainties come from the choice of the DM halo profile.
The signal antiprotons in the cosmic rays can travel far from the solar
system, while the positrons are originated within a few kpc.
Therefore, the prediction of the signal antiproton flux varies with
uncertainties in the halo profile \cite{Bergstrom:1999jc,Donato:2003xg}.
This uncertainties depend on the size of diffusive halo, especially $L$.
For large $L ( \simeq 15$~kpc)
the antiproton flux from the neutralino annihilation may be changed
by several tens of percent depending on the choice of halo profile.
However, for moderate value of $L \lesssim 5$~kpc,
the uncertainties of halo profile is negligible
with respect to other ones such as the diffusive halo size.

Finally, we concentrate the issue of discrimination of the signal from
the background. We showed that the wino-like neutralino annihilation
with mass around 2~TeV leads to the signal antiproton flux comparable
to or larger than the background one. In this case, a bump appears in
the antiproton spectrum. However, it might be still difficult to
recognize presence of the bump from the observed antiproton spectrum
in the upcoming experiments such as PAMELA and AMS-02, compared with
cases of the positron flux from the DM annihilation.

The positron flux has a high-energy component, which is produced more
directly from the leptonic weak boson decays, so that the
signal-background ratio for the higher energy positron flux is better.
However, since the spectrum of the signal antiproton flux is
featureless even at high energy, it suffers more from uncertainties of
the background.

The background antiprotons at high-energy mainly come from the
interaction of the primary protons in the cosmic rays as discussed in
the previous section.  The proton spectrum is mainly determined by the
source term, that is, the injection from supernovae.  The high-energy
protons are produced by the shock-wave acceleration.  It implies that
the spectrum shows the power-law behavior in terms of the energy.
However, it is difficult to predict the slope of the power law.  Thus,
the background flux also has an ambiguity in the slope at a
high-energy range. Varying the slope at the source term is immediately
reflected to the slope of the antiproton flux at the earth. As a
result, with lack of the knowledge of the slope of the source term, it
is difficult to distinguish the signal from the background using only
the slope of spectrum.

It may be important to observe the antiprotons with energy around the
neutralino mass so that the bump may be recognized. When the wino mass
is around 300~GeV and the thickness of the diffusion zone $L$ is
large, the whole structure of the bump might be figured out in the
experiments. However, when the mass is around 2~TeV, it would be a
hard job to detect antiprotons with such a high energy.

\section{Conclusions}

In this article, we have studied indirect detection of the wino-like
neutralino DM using cosmic ray positron and antiproton observations.
Non-perturbative effect enhances the neutralino annihilation cross
section when the mass is larger than about 1~TeV. Especially, when the
mass is around 2~TeV, the cross section is enhanced significantly due
to the resonance effect of the bound state, which is  composed of the
wino-like neutralinos and charginos. In those cases, the cosmic ray
positron and antiproton fluxes produced by the neutralino annihilation
in the galactic halo also enhance, and the sensitivities of the
upcoming experiments, such as PAMELA and AMS-02, are improved for the
heavier neutralino DM.  It is noticed that the relic abundance of the
wino-like neutralino in the universe is explained by the thermal
scenario when the mass is around 2~TeV. It might be difficult to study
such a heavy neutralinos in experiments except for observation of
the cosmic rays. Even in the direct DM detection, the sensitivity
should cover $10^{-(46-47)}$cm$^2$ for the spin-independent cross
section so that the heavy wino-like neutralino is detected
\cite{Hisano:2004pv}.  We have concentrated mainly the heavy wino-like
neutralino and have evaluated the positron and antiproton fluxes from
the neutralino annihilation using the diffusion model.

We found that both positron and antiproton fluxes increase
significantly around the resonance ($m\sim2$~TeV). However, the
positron flux measurement has more prospects to detect the heavy
wino-like neutralino DM, compared with the antiproton one.  The signal
positron flux exceeds the expected background for the positron energy larger
than about 100~GeV, and the spectrums in the positron flux and the
positron fraction are significantly deviated from the background
ones. In addition, it is plausible that the signal positron spectrum
at high energy is less sensitive to the astrophysical parameters in
the diffusion model or the DM halo profile, since the positrons we
observe are produced within a few kpc around the solar system. PAMELA
and AMS-02 have good sensitivities in a broad region of the positron
energy 10~GeV $\lesssim E \lesssim$ 270~GeV. Thus, they may
distinguish whether the heavy wino-like neutralino is the DM.

We have also discussed the HEAT anomaly in a positron energy range
4~GeV $\lesssim E \lesssim$ 20~GeV.  The positron flux from the heavy
wino-like neutralino annihilation with mass 2~TeV is consistent with
it within the experimental error.
It is amazing that the wino-like neutralino can explain
both the DM relic abundance and the HEAT anomaly
even when the mass is around 2~TeV.
        
The antiproton flux from the wino-like neutralino annihilation may be
comparable to or larger than the expected background for the mass around
2~TeV. However, it is strongly dependent on the astrophysical
parameters in the diffusion model. In addition to it, it might be
difficult to discriminate the signal from the background,
since the antiproton spectrum is featureless.

\section*{Acknowledgments}

The work was supported in part by a Grant-in-Aid of the Ministry of
Education, Culture, Sports, Science, and Technology, Government of
Japan (No.~13135207 and 14046225 for JH and No. 16081211 for SM).

\section*{Appendix A: Solution of diffusion equation for positron signal}

Here we show how to solve the diffusion equation for the positron
signal from the DM annihilation. With use of dimensionless
parameter $\epsilon = (E/$~1GeV), the equation~(\ref{eq:diffusion}) is
rewritten under the steady state condition as
\begin{equation}
 K(\epsilon)\nabla^2f_{e^+}(\epsilon,r,z)
 +
 \frac{\partial}{\partial \epsilon}
 \left(b(\epsilon)f_{e^+}(\epsilon,r,z)\right)
 +
 Q(\epsilon,r,z)
 =
 0~,
\end{equation}
where $K(\epsilon) = K_0(C + \epsilon^{\alpha})$ and $b(\epsilon) =
\epsilon^2/\tau$.  The values for $K_0$, $C$, $\alpha$, and $\tau$ can
be read off in text.

We use the cylinder coordinate, so the differentiation $\nabla^2$ is
written as $\nabla^2 = \partial^2_r + r^{-1}\partial_r
+ \partial_z^2$. The source term $Q(\epsilon,r,z)$ including the
information of the DM annihilation is
\begin{equation}
Q(\epsilon,r,z) = \frac12
 \left({n(r,z)}\right)^2
 \sum_{f}
 \langle \sigma v \rangle _f
 \left( \frac{dN_{e^+}}{d \epsilon} \right)_f ~,
\end{equation}
where $f$ means the final state of the DM annihilation and $
( dN_{e^+} / d \epsilon )_f $ is the fragmentation function for the final
state $f$.  We impose the boundary condition so that the density of
positron $f_{e^+}(\epsilon,r,z)$ becomes zero at the surface of the
diffusion zone, which is given by a cylinder with radius $R$ and
half-height $L$.

Due to the boundary condition, it is convenient to expand the density by
the zeroth-order Bessel function $J_0$ for the coordinate $r$
and by a sine function for $z$,
\begin{equation}
 f_{e^+}(\epsilon,r,z)
 =
 \sum_{m,n = 1}^\infty
 A_{n,m}(\epsilon)
 J_0\left(\frac{\zeta_n}{R}r\right)
 \sin\left(\frac{m\pi}{2L}(z - L)\right)~,
 \label{eq:dndE}
\end{equation}
where $\zeta_n$ are successive zeros of the function $J_0$. Using the
expansion, it is obvious that the density satisfies the boundary
condition above.

We comment on the some properties of the Bessel function $J_0$
here. It satisfies a following differential equation,
\begin{equation}
 \frac{d^2}{d r^2}J_0\left(\frac{\zeta_n}{R}r\right)
 +
 \frac{1}{r}\frac{d}{dr}J_0\left(\frac{\zeta_n}{R}r\right)
 +
 \frac{\zeta_n^2}{R^2}J_0\left(\frac{\zeta_n}{R}r\right)
 =
 0~,
\end{equation}
and has a following orthogonal relation
\begin{equation}
 \int_0^R rdr J_0\left(\frac{\zeta_i}{R}r\right)
              J_0\left(\frac{\zeta_j}{R}r\right)
 =
 \frac{1}{2}J_1^2(\zeta_i)R^2\delta_{ij}~,
\end{equation}
where $J_1$ is the first-order Bessel function.

Substituting Eq.~(\ref{eq:dndE}) into the diffusion equation
and using the differential equation and orthogonal relation above, we obtain
\begin{equation}
 \frac{d A_{n,m}}{d \epsilon}
 +
 \frac{2}{\epsilon}A_{n,m}
 -
 \left(\frac{\zeta_n^2}{R^2} + \frac{m^2\pi^2}{4L^2}\right)
 \frac{K_0\tau(C + \epsilon^\alpha)}{\epsilon^2}A_{n,m}
 =
 -\frac{\tau}{\epsilon^2}Q_{n,m}(\epsilon)~.
 \label{positron energy diffuction}
\end{equation}
Here, we also expand the source term $Q$ by the Bessel and sine
functions, and the coefficients of the expansion $Q_{n,m}$ are written
\begin{equation}
 Q_{n,m}(\epsilon)
 =
 \frac{2}{J_1^2(\zeta_n)R^2 L}
 \int_0^R rdr\int_{-L}^L dz
 Q(r,z,\epsilon)
 J_0\left(\frac{\zeta_n}{R}r\right)
 \sin\left(\frac{m\pi}{2L}(z - L)\right)~.
\end{equation}

The boundary condition for $A_{n,m}$ is given by
$A_{n,m}(\epsilon_{\rm max}) = 0$, where $\epsilon_{\rm max} \equiv
{\rm max}\left({\rm supp}(Q_{n,m})\right) (\sim m)$.\footnote{
The symbol, ${\rm supp}(f(x))$, implies regions of $x$ in which $f(x)\ne 0$.
} The condition means that $A_{n,m}(\epsilon) = 0$ if $\epsilon \ge
\epsilon_{\rm max}$. The function $A_{n,m}(\epsilon)$ must be a
continuous function of $\epsilon$. Solving Eq.~(\ref{positron energy
  diffuction}), we obtain
\begin{eqnarray}
 A_{n,m}
 &=&
 \int_\epsilon^{\epsilon_{\rm max}}
 d\epsilon'
 \tau Q_{n,m}(\epsilon')
 \frac{1}{\epsilon^2}
 \nonumber \\
 &&\times
 \exp
 \left[
  \left(\frac{\zeta_n^2}{R^2} + \frac{m^2\pi^2}{4L^2}\right)
  K_0\tau
  \left(
   - \frac{C}{\epsilon}
   + \frac{C}{\epsilon'}
   + \frac{\epsilon^{\alpha - 1}}{\alpha - 1}
   - \frac{(\epsilon')^{\alpha - 1}}{\alpha - 1}
  \right)
 \right]~.
\end{eqnarray}
Substituting $A_{n,m}$ into Eq.~(\ref{eq:dndE}), we obtain the number
density $f_{e^+}(\epsilon,r,z)$.

\section*{Appendix B: Solution of diffusion equation for antiproton signal}

The strategy of solving the diffusion equation for the antiproton
signal in  Eq.~(\ref{eq:diff})  is essentially the same as that in the
positron case. In the equation, the term representing the energy loss
of the particles, which has differentiation with respect to $T$,
is absent. It makes it much easier to solve the equation than that of
positrons. Thus, we show only the result here.  For more detailed
calculations, see  Refs.~\cite{Barrau:2001ev,Maurin:2002ua}.

The number density of antiprotons at the solar system,
$f_{\bar p}(T, \vec{r}_\odot)$, is given by
\begin{eqnarray}
 f_{\bar p}(T, \vec{r}_\odot)
 &=&
 \sum_{i = 1}^\infty
 \exp\left(-\frac{V_c L}{2K}\right)
 \frac{y_i(L)}{A_i\sinh(S_iL/2)}
 J_0\left(\zeta_i\frac{r}{R}\right)~,
 \nonumber \\
 y_i(z)
 &=&
 2\int_0^z dz'
 \exp\left(\frac{V_c(z - z')}{2K}\right)
 \sinh\left(\frac{S_i(z - z')}{2}\right)
 Q_i(T, z')~,
 \label{diff2}
\end{eqnarray}
where $Q_i$ is the coefficient of the expansion of the source term
$Q(T,r,z)$ by the Bessel function, 
\begin{equation}
 Q(T,r,z)
 =
 \sum_{i = 1}^\infty
 Q_i(T,z)J_0\left(\zeta_i\frac{r}{R}\right)~.
\end{equation}
The parameter $A_i$ in Eq.~(\ref{diff2}) includes the information
about the propagation of antiprotons, and it is given as
\begin{eqnarray}
 A_i
 =
 2h\Gamma_{\rm inel} + V_c + KS_i\coth\left(\frac{S_i L}{2}\right)~,
 \qquad
 S_i
 =
 \sqrt{\left(\frac{4\zeta_i^2}{R^2} + \frac{V_c^2}{K^2}\right)}~.
\end{eqnarray}
From this solution, we can derive the simple relation between the source term
and the density. Assuming that the source term scales as
$Q\propto E^{-\alpha}$, the number density behaves as
$f_{\bar p}(T, \vec{r}_\odot)\propto E^{-\alpha -\delta}$,
because of $A_i\propto K(E)\propto E^{\delta}$.

\end{document}